\documentclass{nature}

\usepackage{graphicx}
\usepackage{tabularx}
\makeatletter
\let\saved@includegraphics\includegraphics
\AtBeginDocument{\let\includegraphics\saved@includegraphics}
\renewenvironment*{figure}{\@float{figure}}{\end@float}
\makeatother
\usepackage{booktabs}

\usepackage[colorlinks=true,linkcolor=blue,citecolor=blue,urlcolor=blue]{hyperref}
\usepackage{authblk}
\usepackage{amsmath}
\usepackage{bm}
\usepackage{amsfonts,amssymb}
\usepackage{color}
\usepackage{float}
\usepackage{siunitx}
\usepackage{cleveref}
\usepackage{ulem}
\DeclareSIUnit\rydberg{Ry}
\DeclareSIUnit\atomicunit{a.u.}
\DeclareSIUnit\bohr{\text{\ensuremath{a_0}}}

\newcommand{\VEC}[1]{\mathbf{#1}}

\begin{document}

\title{Kalman filter enhanced Adversarial Bayesian optimization for active sampling in inelastic neutron scattering} 

\author[1,2]{Nihad Abuawwad}

\author[3]{ Yixuan Zhang}

\author[1,2,*]{Samir Lounis}

\author[3,*]{Hongbin Zhang}

\affil[1]{Peter Gr\"unberg Institute, Forschungszentrum J\"ulich \& JARA, 52425 J\"ulich, Germany}
\affil[2]{Faculty of Physics, University of Duisburg-Essen and CENIDE, 47053 Duisburg, Germany}
\affil[3]{ Institute of Materials Science, Technical University Darmstadt, Darmstadt 64287, Germany}
\affil[*]{hzhang@tmm.tu-darmstadt.de; s.lounis@fz-juelich.de}

\maketitle

\section*{Abstract}
\begin{abstract}
Spin waves, or magnons, are fundamental excitations in magnetic materials that provide insights into their dynamic properties and interactions. Magnons are the building blocks of magnonics, which offer promising perspectives for data storage, quantum computing, and communication technologies. These excitations are typically measured through inelastic neutron or x-ray scattering techniques, which involve heavy and time-consuming measurements, data processing, and analysis based on various theoretical models. Here, we introduce a machine learning algorithm that integrates adaptive noise reduction and active learning sampling, which enables the restoration from minimal inelastic neutron scattering point data of spin wave information and the accurate extraction of magnetic parameters, including hidden interactions. Our findings, benchmarked against the magnon spectra of CrSBr, significantly enhance the efficiency and accuracy in addressing complex and noisy experimental measurements. This advancement offers a powerful machine learning tool for research in magnonics and spintronics, which can also be extended to other characterization techniques at large facilities.

\end{abstract}

\maketitle

\section*{Introduction}

Collective spin excitations in two-dimensional (2D) materials are a significant area of research in condensed matter physics, primarily due to their potential applications in quantum computing, spintronics, magnonics and other advanced technologies~\cite{Pirro2021,Yuan2022}. These excitations, commonly known as magnons or spin waves, consist of quantized spin oscillations that travel through a material~\cite{doi:10.1126/sciadv.1603150, Chumak2015}. A primary goal in this field has been to characterize wide classes of these excitations, facilitated by advances in spectroscopic techniques such as neutron scattering. These methods assess the kinematics of scattered neutrons to elucidate the dispersion relations, lifetimes, and amplitudes of spin excitations. 
However, neutron scattering faces challenges due to limited neutron sources, lower neutron flux relative to other sources, and minimal neutron scattering cross-sections. Additionally, this experiment is expensive and time-consuming, demanding significant effort to extract the magnon spectra and the underlying complex magnetic interactions.
The analysis of Inelastic Neutron Scattering (INS) data typically requires sophisticated models that account for complex interactions within the studied systems. Theoretical frameworks such as linear spin wave theory (LSWT) \cite{PhysRev.87.568, PhysRev.102.1217}, ab initio calculations \cite{doi:10.1021/acs.inorgchem.6b02312, PhysRevB.105.134304}, and combined approaches like lattice dynamics and quantum Monte Carlo simulations \cite{VAJK200393,PhysRevB.102.134312} are commonly used. However, these comprehensive modeling approaches require substantial computational resources and face significant challenges. For example, capturing realistically all relevant interactions is crucial for precise characterization, yet the complexity of these interactions makes developing accurate predictive models difficult. This complexity leads to discrepancies between model predictions and experimental data, necessitating iterative refinements that are resource-intensive and time-consuming. 

To address the previous challenges in neutron scattering experiments and their related theoretical models, a promising approach involves incorporating machine learning techniques into the planning and prediction stages. Machine learning has already proven effective in other types of experiments by automating data processing, enhancing the accuracy of parameter predictions, and refining experiment design. 
For example, x-ray absorption spectroscopy (XAS) 
benefitted from different algorithms such as Adversarial Bayesian optimization (ABO)~\cite{Zhang2023}, the Radial Basis Functions (RBF)~\cite{Guda2021} leading to significant improvements in the experimental and computational analysis. Also, various featurization techniques were explored to assess their impact on the performance of machine learning models for XAS analysis in both classification and regression tasks~\cite{doi:10.1021/acs.chemmater.3c02584}.
Moreover, techniques such as nonlinear autoencoders streamline the handling of complex datasets, improving experimental setups' efficiency. For example, in studies involving complex systems like spin ice, machine learning enables tuning of Hamiltonian models under varying experimental conditions such as pressure and temperature, leading to improved predictions of material behaviors and phase diagrams~\cite{chen2021integration,springernature2022integration,samarakoon2022integration}. Recently, Convolutional Neural Networks (CNN), trained using linear spin wave simulations, have been employed in inelastic neutron scattering experiments to differentiate between two feasible magnetic exchange models.~\cite{Butler_2021} The performance of CNNs heavily depends on the quality and the size of the training data. However, the data from neutron scattering experiments are noisy and costly to collect, which makes it very difficult to meet the criteria for such a database. 

In our study, we introduce a machine learning algorithm that integrates active learning sampling with linear spin wave theory leading to Kalman Filter enhanced Adversarial Bayesian Optimization (KFABO) algorithm for approximating the magnon spectrum using a minimal number of both sampling points and iterations. With a minimal number of iterations, the algorithm is capable of addressing noisy neutron scattering data, providing reliable magnetic interactions that recover the experimental spectra, and even unlocking hidden or weak interactions such as those induced by spin-orbit coupling.

\begin{figure}[H]
    \centering
    \includegraphics[width=0.99\textwidth]{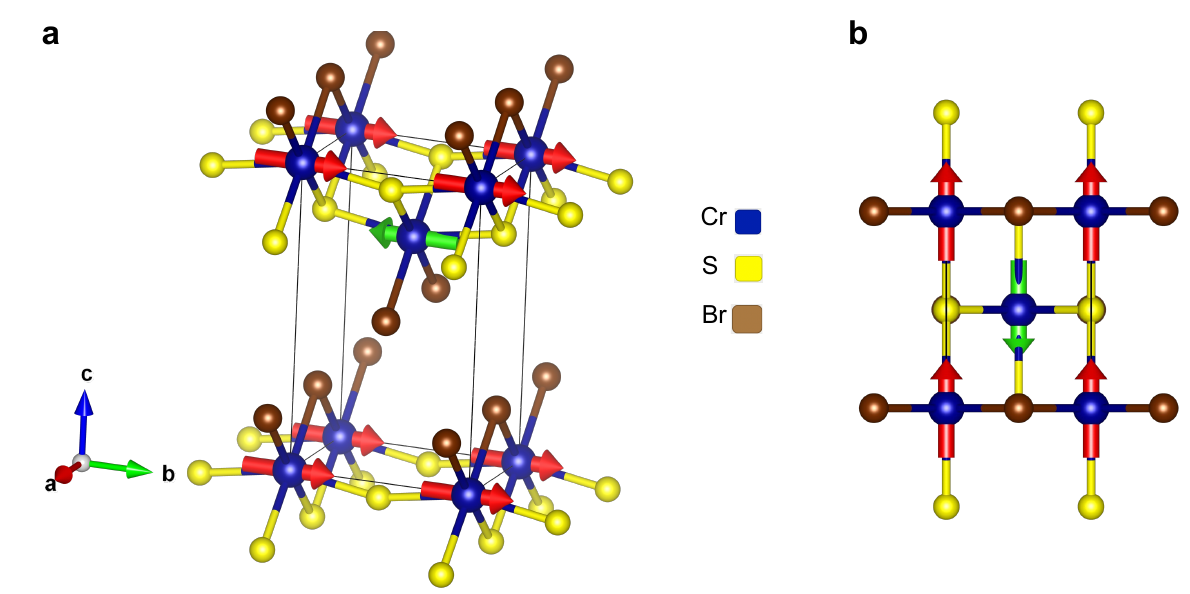}
   \caption{\textbf{Illustration of bulk CrSBr crystal structure.} \textbf{a}, \textbf{b} The side and top views of CrSBr. The magnetic ground state is ferromagnetic within the plane and interlayer antiferromagnetic, which leads to an antiferromagnetic state. }
\label{fig:crystal_structure}
\end{figure}

To corroborate our findings, we explore the antiferromagnetic (AFM) two-dimensional CrSBr material investigated by neutron scattering experiments in Ref.~\cite{https://doi.org/10.1002/advs.202202467}. CrSBr is particularly intriguing due to its unique magnetic properties such as a strong spin-orbit coupling imposing an in-plane magnetization and a large N\'eel temperature ($T_N$) of 132 K~\cite{doi:10.1021/acs.nanolett.4c00624, Goser1990, LopezPaz2022, Telford2020}, which promotes this material for advanced spintronic applications. In its bulk form, the individual layers are ferromagnetic (FM), while they couple among each other in an AFM fashion as shown in Figure~\ref{fig:crystal_structure}. The reported experimental spectra~\cite{https://doi.org/10.1002/advs.202202467} are significantly noisy, which provides an ideal test case for our algorithm. Moreover, the same work provides a fit of the experimental spectra enabling the extraction of Heisenberg exchange interactions and Dzyaloshinskii-Moriya interactions but without resolving the interlayer interaction responsible for the AFM behavior of CrSBr.  
Noting that previous ab initio simulations did not recover the associated large N\'eel temperature by predicting negligible interlayer coupling~\cite{Butler_2021}, our algorithm, however, maps from the same and rather noisy experimentally data a significant AFM interlayer coupling, which incidentally is confirmed by our first-principles calculations. 

\section*{Results}

\subsection{Algorithm benchmarking.} 

\begin{figure}
    \centering
    \includegraphics[width=0.90\textwidth]{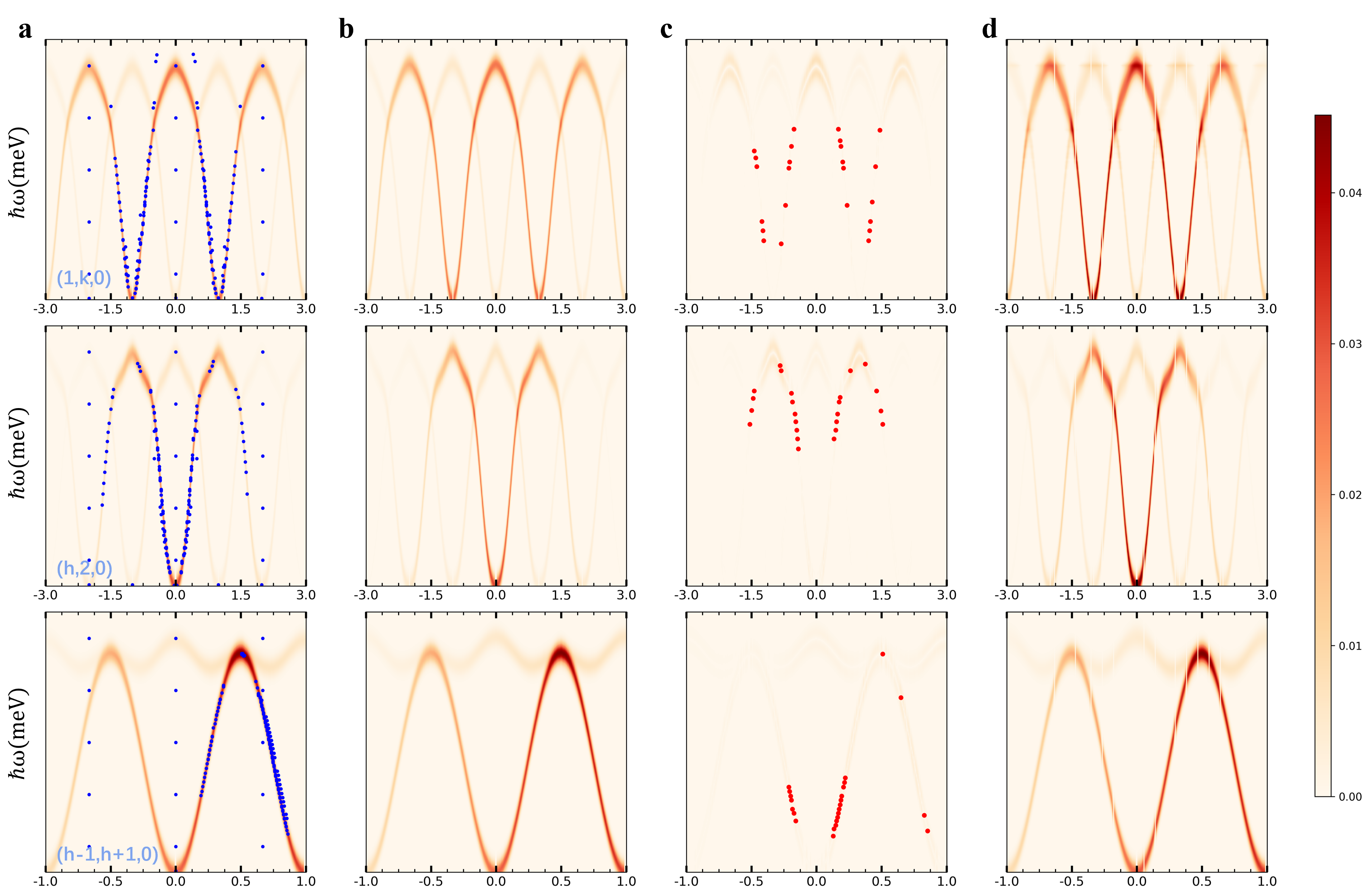} 
    \caption{\textbf{Magnon spectrum for CrSBr with only $J$-values along three q-paths ( \( (1, k, 0) \), \( (h, 2, 0) \), and \( (h - 1, h + 1, 0) \).} \textbf{a} The target magnon spectrum (orange curve) along the three q-paths using the Ref.~\cite{https://doi.org/10.1002/advs.202202467} fitted parameters. The blue points are the active sampling points from previous iterations that were suggested by the sBO. \textbf{b} The fBO fitted the magnon spectrum among the three q-paths using only the information from blue sample points. \textbf{c} The absolute intensity deviation between the standard function and fBO’s prediction, while the red points denote the sample points to be measured for the next round suggested by sBO. \textbf{d} The current state as perceived by the KFABO, which the model samples based on that state. The magnitudes of intensity are described by the color bar.}
    \label{fig:theory-spinw}
\end{figure}

In this section, we combine the LSWT and active-learning sampling together, leading to a KFABO algorithm. The objective of this algorithm is to approximate the calculated magnon spectrum using the minimum number of sampling points necessary. A physical model of LSWT with 8 independent nearest neighboring Heisenberg exchange interactions $J$, feeding the Hamiltonian $- \sum_{i,j} J_{ij} \VEC{S}_i \cdot \VEC{S}_j$ with $\VEC{S}_i$ being the unit vector of the atomic moment at site $i$ (see Method section).
The interactions are injected as parameters in the 
fitting Bayesian optimization (fBO) algorithm. A binning parameter ($dE$)~\cite{Toth_2015} leads to a Gaussian broadening of spin waves spectra.  
This effective broadening is induced by both experimental conditions, electronic and dissipation mechanisms, which limit the lifetime of the magnonic modes. The simulation of the latter requires complex theoretical frameworks, which incorporate time-dependent phenomena and many-body physics~\cite{Costa2004,Buczek2009,Sasioglu2010,Lounis2010, Lounis2015,Dias2015, GORNI2022108500, Ke2021, Poelchen2023}. Both the magnetic interactions and the spectral broadening are automatically fitted by our algorithm during the active learning (AL) process.

As a starter, we use the Heisenberg magnetic exchange interactions (up to eight neighboring interactions) that were extracted from the fit of the experimental data published in Ref.~\cite{https://doi.org/10.1002/advs.202202467} to obtain 
the target ground truth spin wave spectra along three wavevector paths -  \( (1, k, 0) \),  \( (h, 2, 0) \), and \( (h - 1, h + 1, 0) \)- (see  Supplementary Figure 4(c)).
The necessity of fitting all paths together arises from the incompleteness of information provided by individual paths. The detailed analysis in Supplementary (Figure 2  and Table 1) demonstrates the significant differences and potential inaccuracies when fewer paths are used. As aforementioned, since the AFM interlayer interaction was not resolved in the fit reported in Ref.~\cite{https://doi.org/10.1002/advs.202202467}, we consider it to be zero in the first part of our study devoted to the benchmarking of our algorithm. Regarding the broadening, we note that we assume a value of 3 meV for the energy binning parameter.

The algorithm quickly recovers the shape of the spectra in 3 iterations, using only 261 sample data points in total. It precisely predicts the Heisenberg exchange parameters in 8 iterations using 621 data points (cf. Figure~\ref{fig:theory-spinw} and Supplementary image file "Theoretical\_SPINW\_woDMI.gif")~\cite{gifsurl}. The location of these points is depicted in Figure~\ref{fig:theory-spinw}(a), together with the target magnon spectrum (orange curves).
Figures~\ref{fig:theory-spinw}(b) show the calculated magnon spectrum (orange curves) based on the fBO fitted magnetic interaction parameters (cf. Table~\ref{table-1}), with the pixel-by-pixel absolute deviation between the fitted spin wave and the target one are shown in Figures~\ref{fig:theory-spinw}(c). The fitting is quite good, with an average deviation of 0.00013 and a maximum deviation smaller than 0.004 (compared with the normalized maximum peak intensity with a value of 1). The maximum deviation is primarily due to the energy broadening parameter, which can be reduced by further sampling and fitting. 
This parameter quickly converges to a rough range at the very beginning of the algorithm, as it is crucial for determining the intensity dispersion during this period. Subsequently, the algorithm shifts its focus to refining the magnetic parameters rather than this parameter. However, perfectly fitting the broadening parameter requires more points on the fringes of the intensity peaks, which are less informative for refining the magnetic interaction parameters and the shape of the spin wave.

Figure~\ref{fig:theory-spinw}(d) represents the current state as perceived by the algorithm. This state is identified as the optimal state to obtain the most effective information through further sampling, using the fBO fitting results and the latest sampling data. The sampling results indicate that most points are close to the peaks, except for four less effective sampling points, located near the top center of the q-path $(1, k, 0)$. These points were sampled during the fourth iteration and showed about a $10 \text{meV}$ difference from the actual peaks. These points provide crucial information that helps improve the state quality and reduce uncertainty (cf. Supplementary image file "Theoretical\_SPINW\_woDMI.gif", iteration 4, Figure(d))~\cite{gifsurl}.
From this GIF file, it is evident that the AL process converges after eight iterations, and then the information gain progressively diminishes. This convergence can be confirmed by tracking the state changes across iterations. This indicates that state tracking can serve as a powerful tool for determining the stopping criterion. For example, the refinement of the broadening parameter requires more sampling on the fringes of the peaks, which occurs after the state changes converge. Therefore, it would be more effective to stop after convergence. 
What is particularly interesting is that, even though the importance of the parameters (refers to how much each parameter contributes to a model’s prediction) is not explicitly coded into our KFABO algorithm, it can still automatically make decisions as the AL iterations proceed, this confirms the adaptive nature of the model.

Table~\ref{table-1} illustrates the effectiveness of the fitting procedure by comparing our fitted Heisenberg exchange parameters with their corresponding target values while quantifying the disparities through absolute differences. In this case, all three paths are used in combination. The fitting shows consistent accuracy, with absolute differences below 0.05 meV for all J neighbors. For example, the interactions up to the third nearest neighbors $J_1$, $J_2$, and $J_3$ show absolute differences of 0.0373 meV, 0.0306 meV and 0.0043 meV, respectively, indicating a strong correlation, and a good agreement with the ones reported in Ref.~\cite{https://doi.org/10.1002/advs.202202467}. 

Besides the Heisenberg exchange interactions, a fit of the DMI, $- \sum_{i,j} \VEC{D}_{ij}\cdot(\VEC{S}_i \times \VEC{S}_j)$, was extracted in Ref.~\cite{https://doi.org/10.1002/advs.202202467}. Starting from spin waves spectra generated by both interactions, Heisenberg and DMI, 
the KFABO algorithm recovers the shape of the associated magnon spectrum (cf. Supplementary Figure 3) in the third iteration using 261 data points, and it converges in the ninth iteration using 693 data points (cf. Supplementary image file "Theoretical\_SPINW\_wDMI.gif)~\cite{gifsurl} where the parametric accuracy of the final fBO fitted model is only slightly reduced (cf. Supplementary Table 2) compared to the case where DMI is excluded. Unlike the test without DMI, the Sampling Bayesian Optimization (sBO)  sampling is less stable and efficient, with more sampling points deviating from the actual peak positions. This slight reduction in accuracy can be attributed to the addition of DMI, which increases the difficulty in the magnon spectra predictions by fBO. By introducing an extra dimension of DMI into the Hilbert surface of this fitting problem, the complexity and non-convexity of the surface is increased. A detailed discussion on this matter can be found in Supplementary Note 1.

\begin{table}
\centering
\caption{\label{table-1}Comparison between the KFABO fitted and target Heisenberg exchange interactions up to eight nearest neighbors. The target interactions were reported in Ref.~\cite{https://doi.org/10.1002/advs.202202467} after a fit of the experimental data. The table lists the absolute differences.}
\begin{tabular}{|c|c|c|c|}
\hline
\textbf{$J$ neighbours} & \textbf{KFABO-fitted $J$} & \textbf{Target $J$} & \textbf{Absolute difference} \\
 &  \textbf{(meV)} & \textbf{(meV)} & \textbf{(meV)}\\
\hline
1 & -1.9407 & -1.9034 & 0.0373 \\
2 & -3.3426 & -3.3792 & 0.0366 \\
3 & -1.6741 & -1.6698 & 0.0043 \\
4 & -0.1112 & -0.0933 & 0.0179 \\
5 & -0.0805 & -0.0896 & 0.0091 \\
6 & 0.0001 & 0.0000 & 0.0001 \\
7 & 0.3960 & 0.3665 & 0.0295 \\
8 & -0.2881 & -0.2932 & 0.0051 \\
\hline
\end{tabular}
\end{table}

\subsection{Experimental spin wave spectrum fitting. }

\begin{figure}[H]
    \centering
    \includegraphics[width=0.85\textwidth]{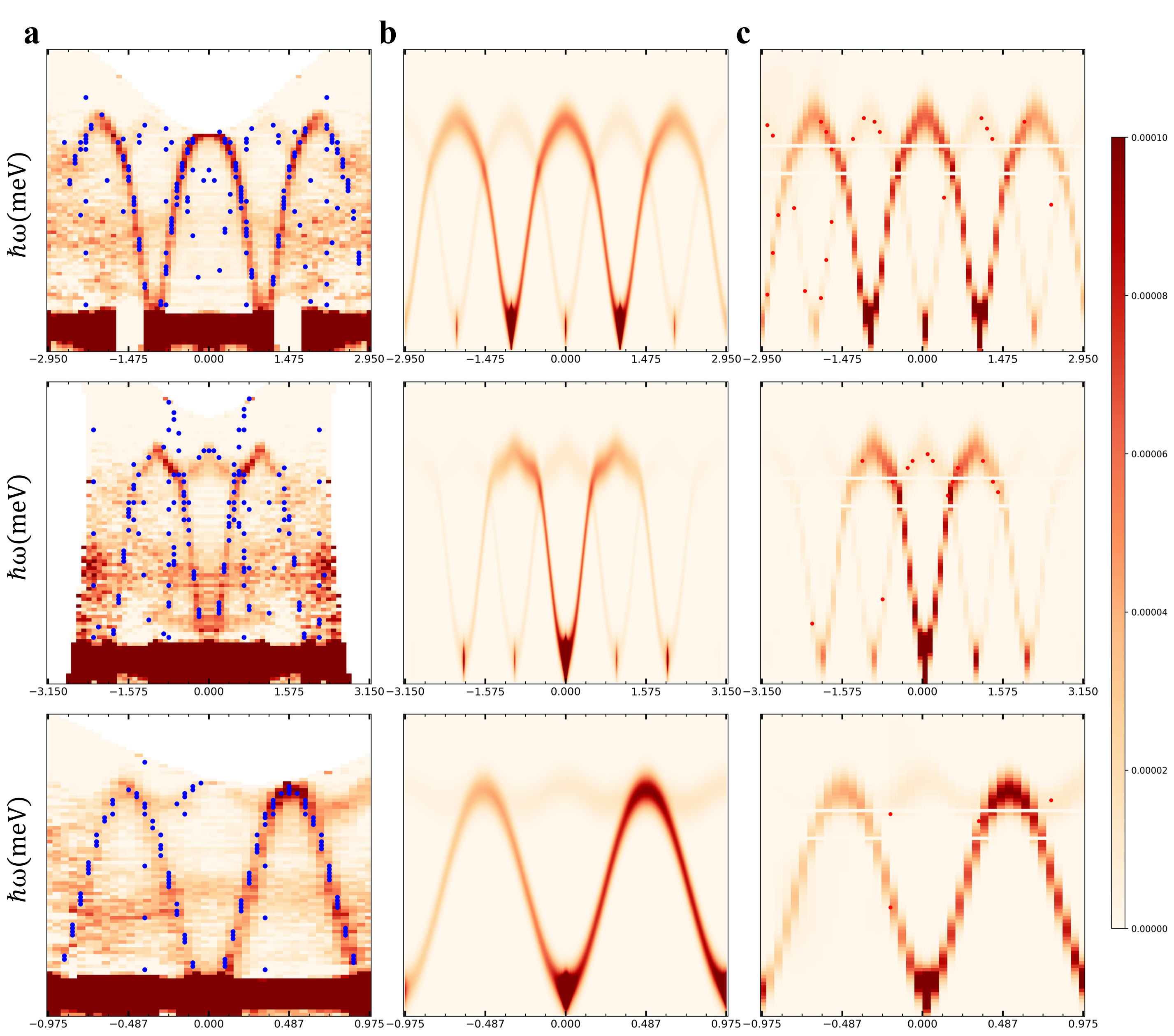} 
    \caption{\textbf{Experimental magnon spectrum for CrSBr along three q-paths ( \( (1, k, 0) \), \( (h, 2, 0) \), and \( (h - 1, h + 1, 0) \).} \textbf{a} Fitting the experimental magnon spectrum (orange curve) along the three q-paths, where the blue points are the sampling points that obtained using the KFABO algorithm. \textbf{b} The calculated spin wave curve 
    along the three q-paths by considering the extracted magnetic parameters. \textbf{c} The current state as perceived by the KFABO which the model samples based on that state, and the red points denote the samples to be measured for the next round suggested by sBO.}
    \label{fig:exp-spinw}
\end{figure}

In this part, we applied the KFABO algorithm to directly fit the experimental spin wave spectrum of CrSBr~\cite{https://doi.org/10.1002/advs.202202467} along three q-paths ( \( (1, k, 0) \), \( (h, 2, 0) \), and \( (h - 1, h + 1, 0) \). As shown in Figure~\ref{fig:exp-spinw}(a), experimental spin wave spectra often include random noise due to various sources such as instrumental limitations, environmental factors, or inherent variability in the material properties. The Kalman filter processes all available measurements to estimate the variables of interest with more accuracy than it would be possible by using a single measurement alone and reducing noise in data.
Figure~\ref{fig:exp-spinw}(a) depicts the experimental magnon spectra along three different q-paths. The intensity of the magnon excitations is shown using a color scale, where darker regions indicate higher intensity. The blue points superimposed on these spectra represent the sample points chosen by the KFABO algorithm during the fitting process. These points are critical as they guide the optimization process. Figure\ref{fig:exp-spinw}(b) shows the calculated magnon spectrum using the fitted magnetic interactions by our algorithm after the noise reduction. A high degree of agreement between the algorithm-fitted spectrum and the original experimental spectrum can be observed. Additionally, by examining the sampling distribution in Figure \ref{fig:exp-spinw}(a), it is evident that the algorithm samples very efficiently despite the strong noise in the experimental data. Most sampling points are concentrated near the peaks, and when sampling noisy regions, the algorithm requires only a small number of sample points to determine the noise level and avoid the associated regions in subsequent iterations (for example, along the q-path $(h,2,0)$ around $h = 2.2$ with an energy of 20 meV).
The algorithm recovers the shape of the spin wave in 3 iterations using 281 data points (cf. Supplementary file "Experimental\_SPINW\_wDMI.gif")~\cite{gifsurl} and it converges after the seventh iteration with 481 data points, as evidenced by the algorithm-perceived state shown in Figure~\ref{fig:exp-spinw}(c).

The parameters fitted by the KFABO algorithm are in a good agreement with the experimentally fitted parameters ~\cite{https://doi.org/10.1002/advs.202202467} with the exception of a small but finite AFM interlayer coupling (6th neighboring magnetic interaction) value of 0.25 meV, see Table~\ref{exp-fit}. Previously, the latter was not resolved with a conventional fitting procedure~\cite{https://doi.org/10.1002/advs.202202467} and was predicted to be negligible by first-principles simulations~\cite{Bo_2023}, which would not explain the large N\'eel temperature characterizing CrSBr.

Motivated by this finding, we proceeded to ab initio simulations of the magnetic properties of CrSBr (see Method section). We extracted the distance-dependent magnetic interactions, presented in Table~\ref{exp-fit} and the Supplementary Figure 4(a). We recover the intralayer FM coupling (cf. Supplementary Figure 4(b)) and predict an in-plane orientation of the magnetic moments due to a magnetic anisotropy energy of 0.15 meV. The algorithm fitted intralayer parameters agree with those obtained from our ab initio simulation as shown in Table \ref{exp-fit}. The smallest absolute differences between the ab initio and the KFABO fitted heseisnberg exchange interactions are found for the the fourth, fifth, seventh and eighth nearest neighboring coupling. The rest of intralayer interactions exhibit a slightly larger absolute differences and remain within an acceptable consistency range.
It is worth mentioning that due to the complexity of the experimental procedure and the non-uniform distribution of experimental noises in the magnon spectra measurements, deviations from the independent first-principles results are expected. Given this context and the fact that the algorithm can only access the measured points, it is impressive that the algorithm achieves such comparable parameters estimates. More importantly, the first-principles simulations recover the interlayer AFM coupling of 0.21 meV, which leads to a N\'eel temperature of 130 K as verified by our Mont\'e Carlo simulations.

Moreover, we calculated the DMI and found a finite value 0.31 meV only for the first nearest neighbors, which is in excellent agreement to the one extracted from the experimental data (0.35 meV).
This highlights the precision that KFABO can achieve in assessing intricate spin-orbit coupling effects. Such precision is crucial for understanding asymmetric energy landscapes in magnetic materials, which in turn can influence domain wall dynamics and skyrmion stability
~\cite{Roessler2006,Costa2010,Fert2017,Nagaosa2013,Chen2013,Emori2013,dosSantos2020,dosSantosDias2023}.
The significant agreement between the fitted parameters and the ab initio ones, as well as between the excitation dispersion spectra demonstrate the effectiveness of the KFABO algorithm in extracting accurate magnetic parameters and capturing the fundamental features of the magnon spectrum from noisy experimental data. This capability is crucial for effectively reducing the number of experimental measurements and accelerating the associated complex processes.

\begin{table}
\centering
\caption{\label{exp-fit} Comparison between the KFABO fitted and ab initio Heisenberg exchange interactions up to
eight nearest neighbors. The absolute differences are provided. The DMI interaction is significant for the nearest neighboring one and reaches a value of 0.35 meV, which agrees well from the one extracted from ab initio (0.31 meV).}
\begin{tabular}{|c|c|c|c|}
\hline
\textbf{$J$ \textbf{neighbors}} & \textbf{KFABO-fitted $J$} & \textbf{Ab initio $J$} & \textbf{Absolute differences} \\ 
 &  \textbf{(meV)} & \textbf{(meV)} & \textbf{(meV)}\\
\hline
1 & -2.29 & -1.90 & 0.39 \\
2 & -3.23 & -3.38 & 0.15 \\
3 & -1.47 & -1.67 & 0.20 \\
4 & -0.16 & -0.09 & 0.07 \\
5 & -0.14 & -0.09 & 0.05 \\
6 & 0.25 & 0.21 & 0.04\\ 
7 & 0.36 & 0.37 & 0.01 \\
8 & -0.24 & -0.29 & 0.05 \\ \hline
\end{tabular}
\end{table}

\section*{Discussion}

In this study, we exposed our approach and results for fitting the spin wave spectra of CrSBr through the KFABO algorithm. First, by combining linear spin wave theory with active-learning sampling, we implemented the KFABO algorithm to approximate the target magnon spectra using minimal sampling points. Our initial fitting uses eight independent Heisenberg exchange parameters and the results, compared to LSWT-based fits reported in Ref.~\cite{https://doi.org/10.1002/advs.202202467}, show that the KFABO algorithm accurately reproduce the magnon spectrum across all q-paths. 
The sampling points that deviate from the peaks have proven to be the informative points, helping the algorithm to converge more quickly and greatly improve the sampling efficiency based on the algorithm perceived state changes tracking. 
Also, the same level of efficiency for sampling can be observed when including the DMI. 

Additionally, the analysis of fitting accuracy using different path configurations reveal that the information provided by some paths may be incomplete, necessitating careful selection of multiple paths to avoid introducing non-convexity(cf. Supplementary Figure 2). To address this issue, we propose to determine the ground state of the material first, and then strategically select q-paths close to this state for detailed analysis. This solution is predicated on the assumption that paths near the ground state provide a more complete and representative magnon spectrum. To validate the previous solution and demonstrate the wide reliability of our algorithm, we applied the algorithm to the complete magnon spectrum of another material with a well-understood ground state, La$_2$CuO$_4$ as shown in the Supplementary Figure 5. By focusing sampling points around the known minimum ground state, it is possible to enhance both the accuracy and comprehensiveness of the magnon spectrum analysis. This strategy not only minimizes the discrepancies in key parameters but also ensures a richer dataset for validating theoretical models.

Regarding the fit of the noisy experimental data, we apply the KFABO algorithm to automatically address and reduce noise in the data. This approach allows for more accurate parameter estimation and very effective sample points, as presented in the experimental magnon spectra along three q-paths. Crucially, the algorithm is capable of quantifying a finite AFM interlayer coupling of 0.25 meV, which was unresolved in the fitting procedure applied in Ref.~\cite{https://doi.org/10.1002/advs.202202467}. Impressively, this value agrees with our ab initio simulations, which together with the rest of distant-dependent magnetic interactions predict a N\'eel temperature of 130K that matches the experimental value~\cite{doi:10.1021/acs.nanolett.4c00624}.

Overall, the good performance of the proposed scheme, which combines machine learning with physical modeling, suggests a promising avenue for future research in material science, especially in areas with significant experimental limitations. The methods developed here have clear potential to efficiently reduce costs and time-consuming numerical and experimental processes, opening new vistas in addressing the physics of magnonics and spintronics.

\begin{methods}
\label{sec:methods}

\subsection{Kalman filter enhanced Adversarial Bayesian Optimization algorithm.}

We combine Linear Spin Wave Theory (LSWT), Active Learning Sampling, and Adaptive Noise Reduction to form the Kalman Filter enhanced Adversarial Bayesian Optimization Algorithm (KFABO). This algorithm integrates two coupled Bayesian Optimization (BO) algorithms with a Kalman filter. The first BO algorithm, termed fBO, employs trust region Bayesian optimization (Turbo) on our linear response model to search for optimal parameters, aiming to minimize the difference between theoretically predicted and real LSW function values of the measured sample points. The second algorithm, termed sBO, is a standard BO that selects sampling points with maximum information gain relative to the current state, specifically, those points that can better characterize the real LSW function given the current samples and the fitted LSW function. Detailed explanations of Turbo and BO can be found in Ref. \cite{Zhang2023}. To enhance model fitting performance and mitigate errors due to experimental noise, a Kalman filter is incorporated. This filter infers the noise distribution across the sampling space based on the current state, which contains only partial and potentially unreliable information and reduces sample noise based on this distribution for the subsequent fBO fitting iteration. The mathematical form of KFABO is expressed as:
\begin{equation}
\text{max}_{\text{sBO}}\left( \text{min}_{\text{fBO}} \left(f_{\text{LSW}}(X)-Y'\right) \right)\;,
\end{equation}
where $f_{\text{LSW}}(X)$ denotes the theoretically predicted spin wave intensity of all the measured points $X$ obtained using the LSW function SpinW code, and $Y'$ is the LSW intensity filtered from the real LSW intensity $Y$ of measured points $X$, where $Y$ is obtained either from theoretical simulations (with hidden parameters) or real experimental measurements.
The fBO predicts the possible spin wave intensity $\bm{\mu}_{\text{LSW}}$ and the relative uncertainties $\bm{\sigma}_{\text{LSW}}$ using Turbo and it can be expressed as:
\begin{equation}
\bm{\mu}_{\text{LSW}}, \bm{\sigma}_{\text{LSW}} = \text{Turbo}(f_{\text{LSW}}(X)-Y')\;.
\end{equation}
The sBO selects the next sampling points that are most likely to contain the most important information about the ground-truth spin waves, based on the current predictions from the fBO and the existing sampling conditions. The mathematical expression of sBO is:
\begin{equation}
X_s = \textbf{\textit{Peak}}(\bm{\mu}_{\text{LSW}}(X) + \text{KI}_{\text{AL}} * \bm{\mu}_{\text{AL}}))\;,
\end{equation}
where $\text{KI}_{\text{AL}} = \bm{\sigma}_{\text{LSW}}(\bm{\sigma}_{\text{AL}}+\bm{\sigma}_{\text{LSW}})^{-1}$ is the Kalman improvement for AL process. $\bm{\mu}_{\text{AL}},\bm{\sigma}_{\text{AL}}=\mathbb{G}_{\text{AL}}(X, Y'-\bm{\mu}_{\text{LSW}})$ are the mean and variance of the residual between $Y'$ and $\bm{\mu}_{\text{LSW}}$ predicted by Gaussian $\mathbb{G}_{\text{AL}}$. The $\textbf{\textit{Peak}}$ function is a screening method that filters out the peaks with the highest intensity within a specified cutoff range and sorts the peaks by considering the sample densities within its range. The top points are then selected as the final sample points $X_s$.

The training set $Y'=[Y_{KF}, Y_s]$ consists of the filtered intensity $Y_{KF}$ related to the sample points $X$ from the previous iteration and the newly measured intensity $Y_s$ of the newly selected samples $X_s$. The filtered $Y_{KF}$ is defined as:
\begin{equation}
\begin{split}
    Y_{KF} &= \bm{\mu}_{\text{LSW}}(X) + \text{KI}_{\varepsilon} * \text{res}(X),\\
    \text{KI}_{\varepsilon} &= (\bm{\sigma}_{\text{LSW}}(\bm{\sigma}_{\varepsilon}+\bm{\sigma}_{\text{LSW}})^{-1},\\
    \text{res}(X) &= \bm{\mu}_{\varepsilon}(X)+r_{\text{trust}}*(Y-\bm{\mu}_{\varepsilon}(X))\;,
\end{split}
\end{equation}
where $\text{KI}_{\varepsilon}$ and $\text{res}(X)$ are the Kalman improvement of the noise reduction process and residual used for the Kalman filtering process. Both variables are estimated using $\bm{\mu}_{\varepsilon},\bm{\sigma}_{\varepsilon}=\mathbb{G}_{\varepsilon}(X, Y-\bm{\mu}_{\text{LSW}})$, which are the predicted mean and variance of residual between the measurement $Y$ and \text{LSW} model prediction $\bm{\mu}_{\text{LSW}}$. The $r_{\text{trust}} = 1-cos(\bm{\mu}_{\varepsilon}(X_s), Y_s-\bm{\mu}_{\text{LSW}}(X))$ is a ratio to evaluate the trustfulness of the Gaussian $\mathbb{G}_{\varepsilon}$. The KFABO workflow is shown in Fig. \ref{fig:KFABO}. 

\begin{figure}
    \centering
    \includegraphics[width=0.9\textwidth]{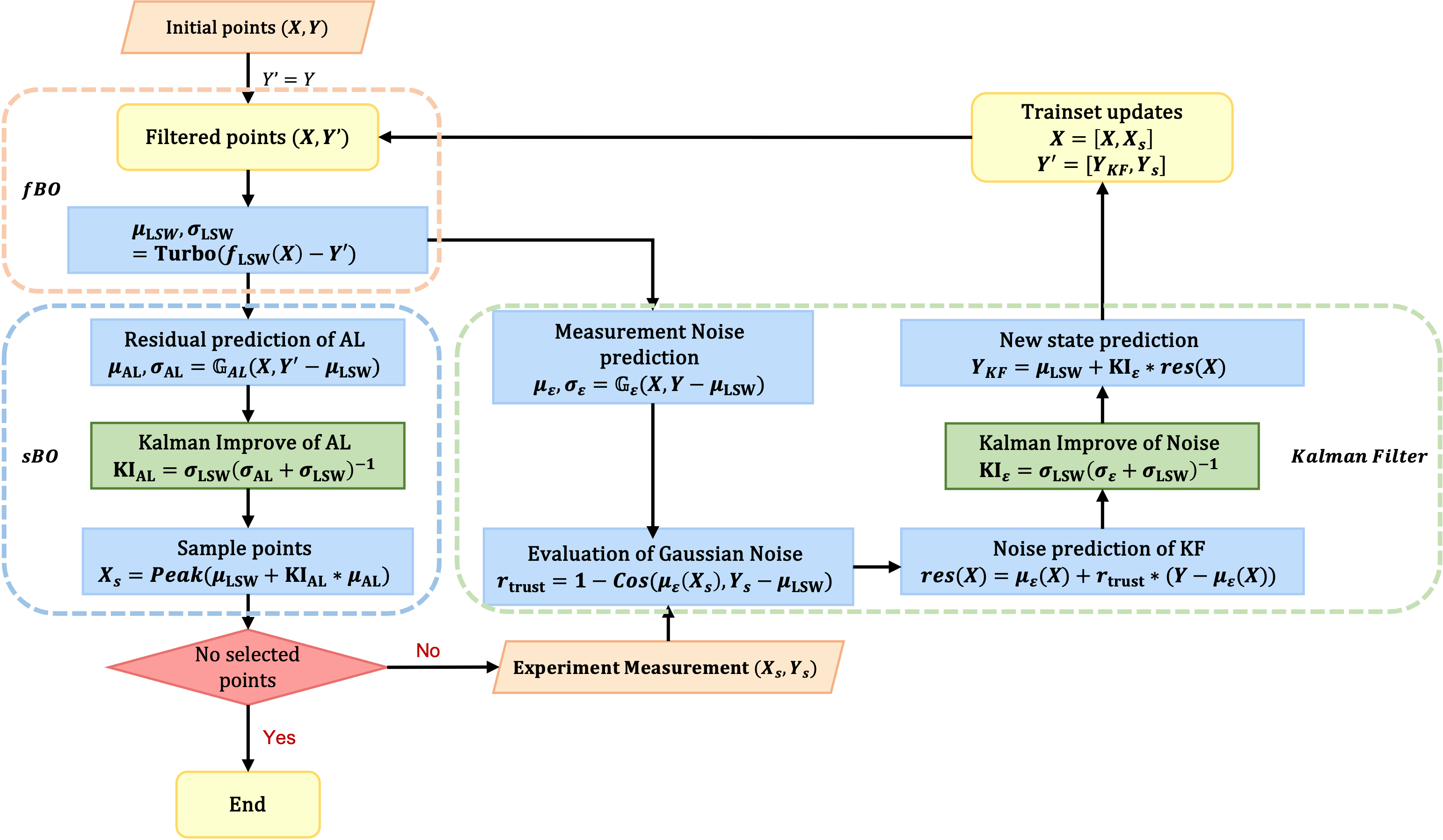} 
    \caption{\textbf{The KFABO workflow.} The fBO is used to search for optimal parameters of the physical model aiming to minimize the difference between theoretically predicted and measured sample points. The sBO is a standard BO that selects sampling points to better characterize the ground truth given the current samples and the fitted physical model. The Kalman filter infers the noise distribution across the sampling space based on the current state and reduces sample noise based on this distribution for the subsequent fBO fitting iteration.}
    \label{fig:KFABO}
\end{figure}

\subsection{First-principles calculations} 

CrSBr consists of 2D layers where magnetic $\mathrm{Cr}^{3+}$ ions are arranged in a rectangular lattice, as depicted in Figure \ref{fig:crystal_structure}(a). First, we relaxed the bulk crystal structure, determining the lattice constants as $a = 3.50\, \text{\AA}$, $b = 4.76\, \text{\AA}$, and $c = 7.96\, \text{\AA}$. The atomic position relaxations were assessed using density functional theory (DFT) as implemented in the Quantum Espresso (QE) computational package \cite{qe} with projector augmented plane wave (PAW) pseudopotentials \cite{ps}.
In our calculations, the generalized gradient approximation (GGA) of Perdew-Burke-Ernzerhof (PBE)\cite{PhysRevLett.77.3865} was used as the exchange and correlation functional.
The plane-wave energy cut-off is \SI{70}{\rydberg}, and the convergence criterion for the total energy is set to \SI{0.01}{\micro\rydberg}.
The self-consistent calculations were performed with a k-mesh of $16 \times 12 \times 8$ points. 
The atomic positions were optimized by ensuring that the residual forces on the relaxed atomic positions were smaller than \SI{1}{\milli\rydberg\per\bohr}.
Once the geometries of bulk CrSBr were established, we explored in detail magnetic properties and interactions with the all-electron full-potential relativistic Korringa-Kohn-Rostoker Green function (KKR-GF) method as implemented in the JuKKR computational package \cite{Papanikolaou2002,Bauer2014, bauer2014development, stefanou1990efficient, russmann2022juktteam}.  
The angular momentum expansion of the Green function was truncated at $\ell_\mathrm{max} = 3$ with a k-mesh of $32 \times 24 \times 16$ points.
The energy integrations were performed including a Fermi-Dirac smearing of \SI{502.78}{\kelvin}, and the local spin-density approximation was employed~\cite{Vosko1980}. 
The magnetic interactions obtained from the first-principles calculations on the basis of the infinitesimal rotation method \cite{inf-rot} are used to parameterize the following classical extended Heisenberg Hamiltonian with unit spins, $|\mathbf{S}| = 1$, which includes the Heisenberg exchange coupling ($J$), the DMI ($D$) and the magnetic anisotropy energy ($K$), with a finer k-mesh of $100\times 80 \times 60$:
\begin{equation}\label{eq:spin_model}
E =  \sum_i K_i (S_i^z)^2
- \sum_{i,j} J_{ij} \VEC{S}_i \cdot \VEC{S}_j - \sum_{i,j} \VEC{D}_{ij}\cdot(\VEC{S}_i \times \VEC{S}_j).
\end{equation}
Here $i$ and $j$ label different magnetic sites within a unit cell. 
Moreover, we have calculated the spin wave curves along with linear spin wave theory (LSWT) as implemented in the SpinW code~\cite{SpinW}, achieving a good agreement with those fitted by the KFABO algorithm as shown in the Supplementary Figure 4(d-f). Finally, the N\'eel temperature was calculated using the Monte Carlo simulation as implemented in Spirit code~\cite{muller2019}.

\end{methods}

\noindent \textbf{Data availability}
The data that support the findings of this study are available from the corresponding author upon reasonable request.
The codes employed for the simulations described within this work are open-source and can be obtained from the respective websites and/or repositories.
Quantum Espresso can be found at \cite{QEurl}. The J\"ulich-developed codes JuKKR is available from \cite{JuKKRurl} while the SpinW code and the Spirit code are downloadable from respectively \cite{SpinW} and \cite{Spiriturl}.

\begin{addendum}

\item This work was supported by the Federal Ministry of Education and Research of Germany in the framework of the Palestinian-German Science Bridge (BMBF grant number 01DH16027).
We acknowledge funding provided by the Priority Programmes SPP 2244 "2D Materials Physics of van der Waals heterobilayer"  (project LO 1659/7-1), SPP 2137 “Skyrmionics” (Projects LO 1659/8-1) of the Deutsche Forschungsgemeinschaft (DFG), and Deutsche Forschungsgemeinschaft (DFG, German Research Foundation) - Project-ID 405553726 - TRR 270.
We acknowledge the computing time granted by the Simulations were performed with computing resources granted by RWTH Aachen University under project p0020362, 
and the computing time provided to them on the high-performance computer Lichtenberg at the NHR Centers NHR4CES at TU Darmstadt. This is funded by the Federal Ministry of Education and Research, and the state governments participating on the basis of the resolutions of the GWK for national high performance computing at universities (www.nhr-verein.de/unsere-partner).

\item[Competing interests] The authors declare no competing interests.

\item[Author contributions] N.A. and Y.Z. performed equal contributions to all the simulations, carried out the initial analysis, and wrote the
the initial draft of the paper, which subsequently benefited from significant input from S.L. and H.Z. H.Z. and S.L. conceived, secured funding for, and supervised the project.

\end{addendum}

\section*{References}
\bibliographystyle{naturemag}
\bibliography{Ref.bib}

\end{document}


\title{Kalman Filter enhanced Adversarial Bayesian optimization for active sampling in inelastic neutron spectroscopy}

\author[1,2,3]{Nihad Abuawwad}

\author[4]{ Yixuan Zhang}

\author[1,2]{Samir Lounis}

\author[4]{Hongbin Zhang}

\affil[1]{Peter Gr\"unberg Institut and Institute for Advanced Simulation, Forschungszentrum J\"ulich \& JARA, 52425 J\"ulich, Germany}
\affil[2]{Faculty of Physics, University of Duisburg-Essen and CENIDE, 47053 Duisburg, Germany}
\affil[3]{Department of Physics, Birzeit University, PO Box 14, Birzeit, Palestine}
\affil[4]{ Institute of Materials Science, Technical University Darmstadt, Darmstadt 64287, Germany}

\maketitle

\begin{figure}[H]
    \centering
    \includegraphics[width=0.99\textwidth]{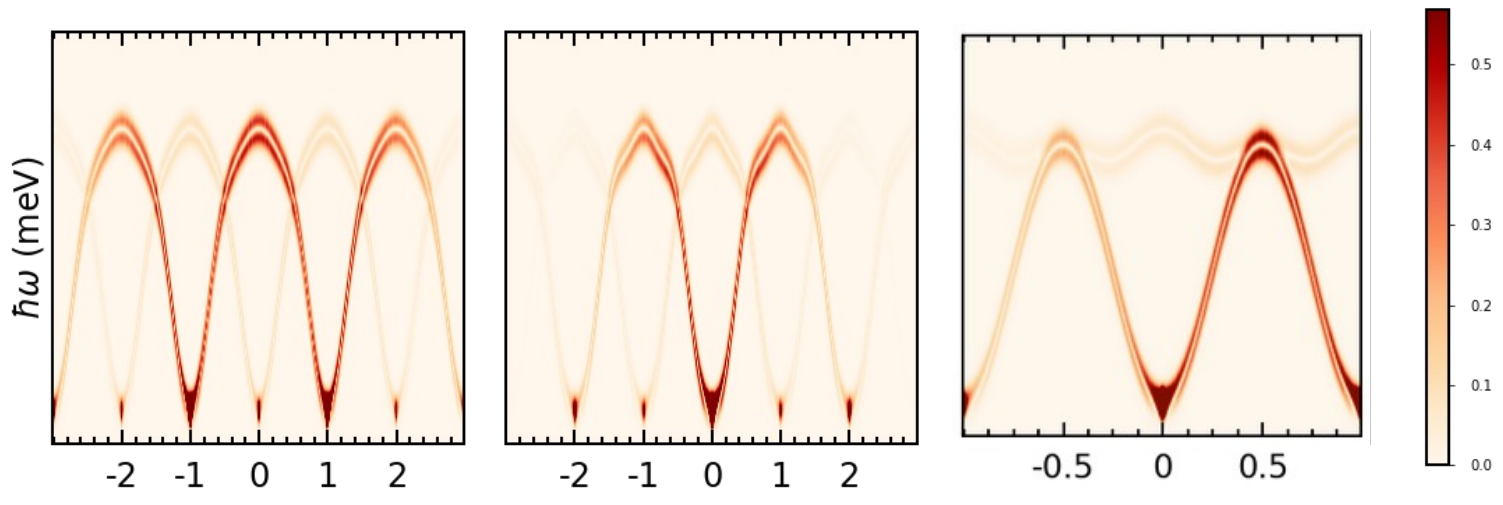}
   \caption{\textbf{The pixel-by-pixel absolute differences plot of bulk CrSBr between the spin wave with and without the interlayer AFM coupling (0.21 meV) for three different wavevector paths: \( (1, k, 0) \), \( (h, 2, 0) \), and \( (h - 1, h + 1, 0) \).} The intensity of two spin waves was normalized into the same scale. The darker the color indicates the larger the differences. The curve with a larger peak for the case of including interlayer AFM coupling.}
\label{fig_0}
\end{figure}

\section{Supplementary Note 1: Algorithm benchmarking with choosing specific q-path, and inclusion DMI.}

In this section, we depict first our fitting using the Heisenberg exchange parameters from Ref. ~\cite{https://doi.org/10.1002/advs.202202467-1} by choosing only a single path \( (1, k, 0) \) as shown in the Supplementary Figure. 2 (a). The KFABO spin wave function aligns reasonably well with the LSWT function, with the maximum and average losses of $0.001278$ and $0.000027$, respectively.
While Supplementary Figure. 2 (b) compares the fitting accuracy when a second path \( (h, 2, 0) \) is added with the maximum and average losses are $0.001407$ and $0.000014$.
Comparatively, the magnetic interactions fitted as shown in Supplementary Table. 1, where one path —\( (1, k, 0) \)- and then two paths —\( (1, k, 0) \), \( (h, 2, 0) \)- are used, the differences of both cases are much higher than the 3-path case, such as $J_1$ showing a significant difference of 0.2593 and 1.7789 in each case. This indicates that the information obtained from these two paths is only part of the information for this fitting problem. If we try to solve the parameter fitting problem by relying only on this incomplete information, there may be many local optimal solutions that are very close to the global optimal solution, which greatly increases the non-convexity of the problem and decreases the probability that we will find the true optimal solution. 

Then we included the DMI to our fitting as shown in  Supplementary Figure 3 and Supplementary image file "Theoretical\_SPINW\_wDMI.gif"~\cite{gifsurl-1}. The KFABO algorithm recovers the shape of the magnon spectrum in the third iteration using 261 data points, and it converges in the ninth iteration using 693 data points where the parametric accuracy of the final fBO fitted model is only slightly reduced (cf. Supplementary Table 2) compared to the case where DMI is excluded. One can see that from Supplementary Table 2, the inclusion of the DMI in parameter fitting slightly diminishes the modeling precision of the KFABO algorithm. Notably, the first, fourth, sixth, and eighth nearest neighboring interactions $J_1$, $J_4$, $J_6$, and $J_8$ (see Supplementary Figure 4a) maintain the same level of accuracy compared to the case where DMI is excluded. While the other interactions still maintain reasonable accuracy but show slightly higher deviations. Unlike the test without DMI, the Sampling Bayesian Optimization (sBO) sampling is less stable and efficient, with more sampling points deviating from the actual peak positions. This slight reduction in accuracy can be attributed to the addition of DMI, which increases the difficulty in the magnon spectra predictions by fBO. By introducing an extra dimension of DMI into the Hilbert surface of this fitting problem, the complexity and non-convexity of the surface are increased.

\begin{figure}[H]
\centering
   \includegraphics[width=0.93\textwidth]{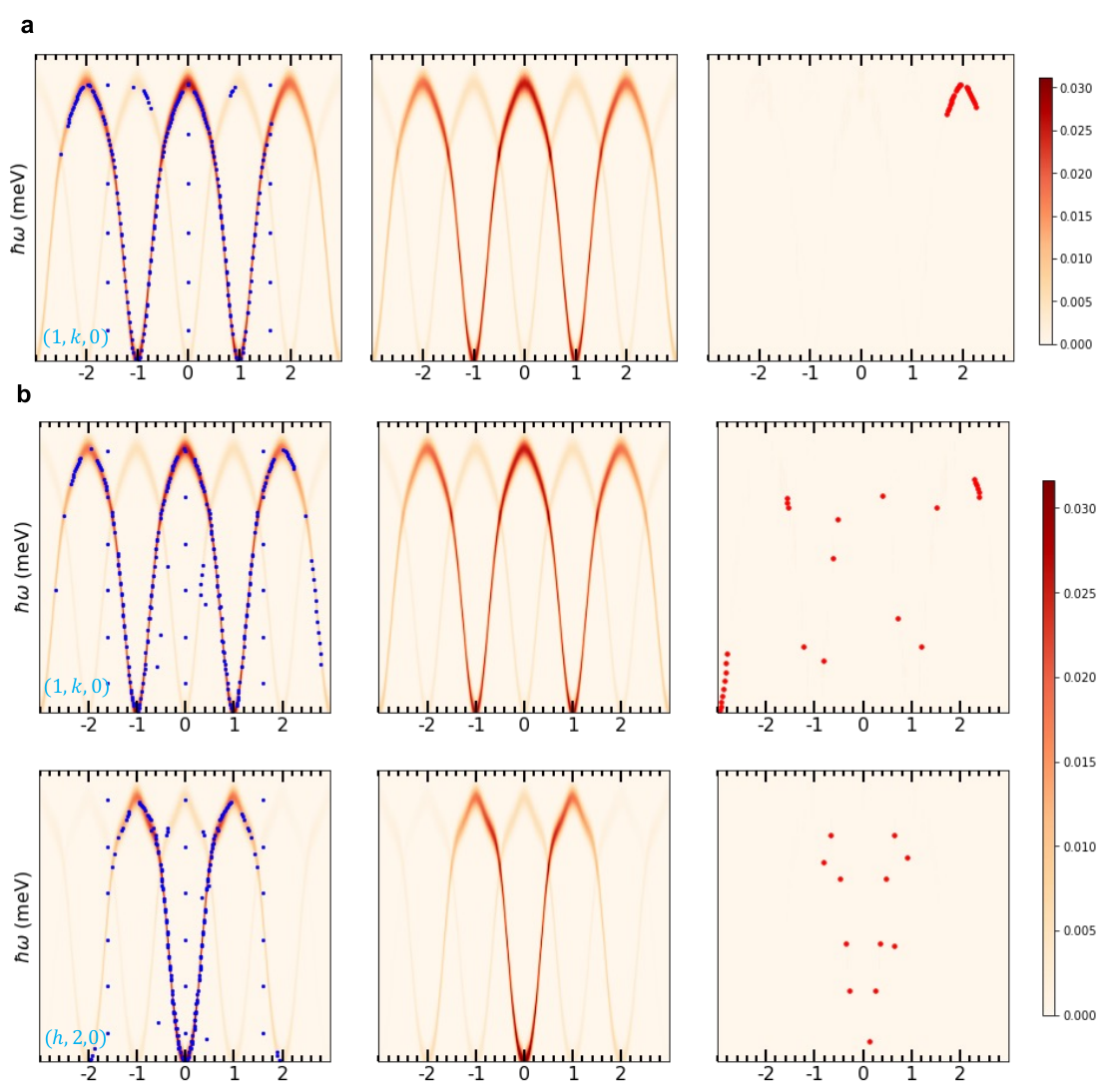}
    \caption{\textbf{Magnon spectrum for CrSBr with only $J$-values for specific q-paths}. \textbf{a} \( (1, k, 0) \). \textbf{b}  \( (1, k, 0) \), and \( (h, 2, 0) \). The left panel represents the target magnon spectrum (orange curve) using the Ref. \cite{https://doi.org/10.1002/advs.202202467-1} fitted parameters, where the blue points are the active sampling points from previous iterations that were suggested by the sBO. The middle panel represents the fBO fitted magnon spectrum using only the information from blue sample points. The last panel represents the absolute intensity deviation between the standard function and fBO’s prediction, while the red points denote the samples to be measured for the next round suggested by sBO. The magnitudes of intensity are described by the color bar.} 
    \label{fig-1}
\end{figure}
\begin{table}
    \centering
    \caption{Comparison of KFABO-fitted parameters versus target parameters using only one \( (1, k, 0) \) path and two-paths  \( (1, k, 0) \) with \( (h, 2, 0) \).The target interactions were reported in Ref.~\cite{https://doi.org/10.1002/advs.202202467-1} after a fit of the experimental data.}
    \label{tab:my_label}
\begin{tabular}{|c|c|c|c|c|}
\hline
\textbf{q-path} & \textbf{$J$ neighbours} & \textbf{KFABO-fitted $J$} & \textbf{Target $J$} & \textbf{Absolute difference} \\  
 &  & \textbf{(meV)} & \textbf{(meV)} & \textbf{(meV)}\\
\hline
\multirow{8}{*}{\( (1, k, 0) \)} & 1 & -1.6441 & -1.9034 & 0.2593 \\
& 2 & -3.3820 & -3.3792 & 0.0028 \\
& 3 & -1.4160 & -1.6698 & 0.2538 \\
& 4 & -0.2207 & -0.0933 & 0.1274 \\
& 5 & -0.0895 & -0.0896 & 0.0001 \\
& 7 & 0.3627 & 0.3665 & 0.0038 \\
& 8 & -0.2898 & -0.2932 & 0.0034 \\
\hline
\textbf{q-path} & \textbf{$J$ neighbours} & \textbf{KFABO-fitted $J$} & \textbf{Target $J$} & \textbf{Absolute difference} \\
 &  & \textbf{(meV)} & \textbf{(meV)} & \textbf{(meV)}\\
\hline
\multirow{8}{*}{\( (1, k, 0) \) and \( (h, 2, 0) \)} & 1 & -0.1245 & -1.9034 & 1.7789 \\
& 2 & -3.0738 & -3.3792 & 0.3054 \\
& 3 & -1.5114 & -1.6698 & 0.1584 \\
& 4 & -0.1752 & -0.0933 & 0.0819 \\
& 5 & -0.3913 & -0.0896 & 0.3017 \\
& 7 & 0.1541 & 0.3665 & 0.2124 \\
& 8 & -0.2920 & -0.2932 & 0.0012 \\
\hline
\end{tabular}
\end{table}

\begin{table}
\centering
\caption{Comparison between the KFABO fitted and target Heisenberg exchange interactions up to
eight nearest neighbors. The absolute differences are provided. The DMI interaction is significant for the nearest neighboring one and reaches a value of 0.2850 meV, which agrees well with the target one (0.3100 meV).}
\label{tab:DMI_parameters}
\begin{tabular}{|c|c|c|c|}
\hline
\textbf{Parameters} & \textbf{KFABO-Fitted} & \textbf{Target parameters} & \textbf{Absolute difference} \\
 &  \textbf{parameters (meV)} & \textbf{(meV)} & \textbf{(meV)}\\
\hline
1 & -1.8772 & -1.9034 & 0.0262 \\
2  & -3.2701 & -3.3792 & 0.1091 \\
3  & -1.7207 & -1.6698 & 0.0509 \\
4  & -0.0729 & -0.0933 & 0.0204 \\
5  & -0.1898 & -0.0896 & 0.1002 \\
6  &  0.0008 &  0.0000 & 0.0008 \\
7  &  0.4719  & 0.3665 & 0.1054 \\
8  & -0.2869 & -0.2932 & 0.0063 \\ \hline
\end{tabular}
\end{table}

\begin{figure}[H]
\centering
   \includegraphics[width=0.92\textwidth]{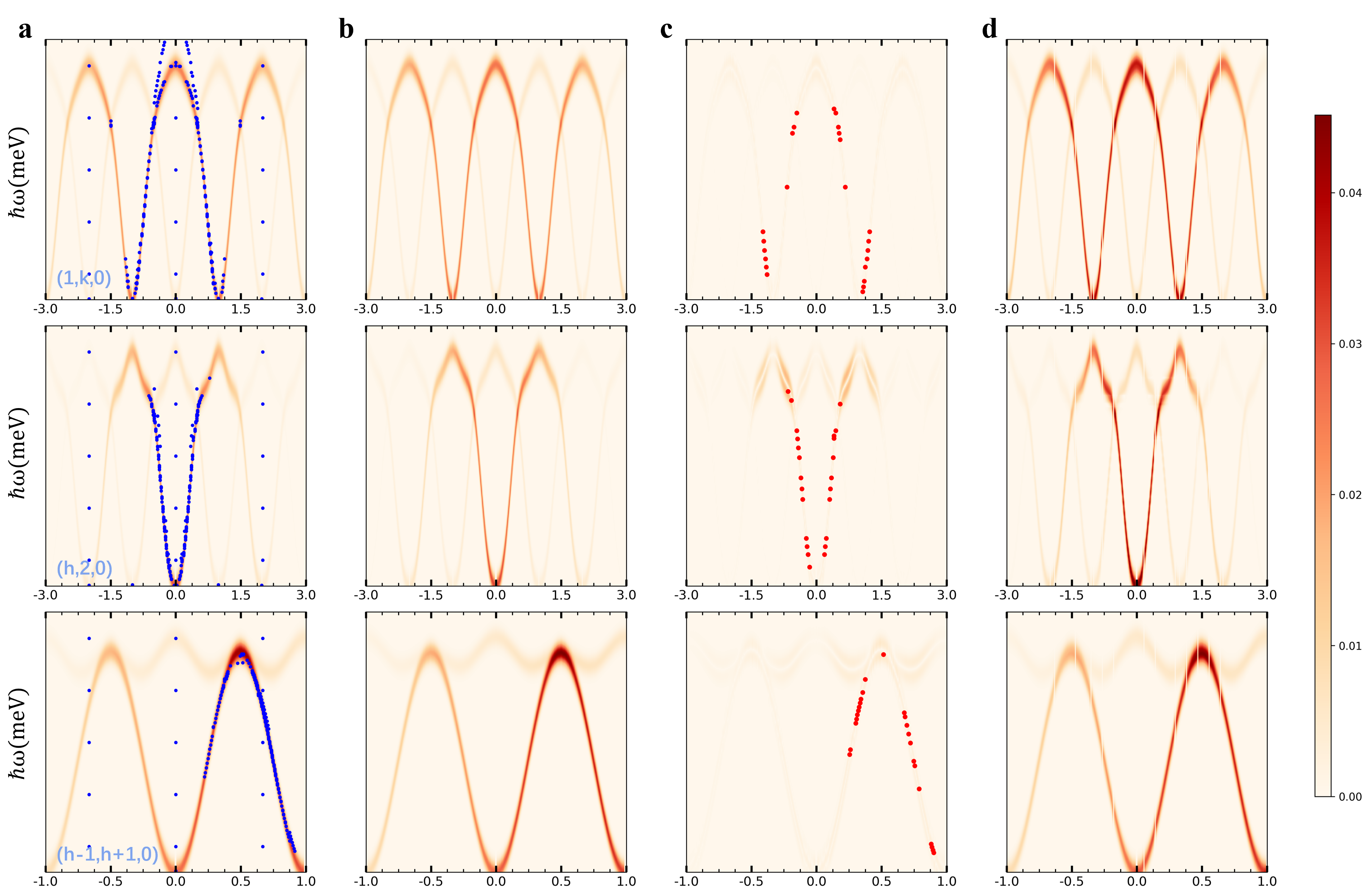}
        \caption{\textbf{Magnon spectrum for CrSBr with including the DMI along three q-paths ( \( (1, k, 0) \), \( (h, 2, 0) \), and \( (h - 1, h + 1, 0) \).} \textbf{a} The target magnon spectrum (orange curve) along the three q-paths using the Ref. \cite{https://doi.org/10.1002/advs.202202467-1} fitted parameters. The blue points are the active sampling points from previous iterations that were suggested by the sBO. \textbf{b} The fBO fitted the magnon spectrum among the three q-paths using only the information from blue sample points. \textbf{c} The absolute intensity deviation between the standard function and fBO’s prediction, and the red points denote the samples to be measured for the next round suggested by sBO. \textbf{d} The current state as perceived by the KFABO, which the model samples based on that state. The magnitudes of intensity are described by the color bar.}
    \label{fig-1}
\end{figure}

\begin{figure}[H]
    \centering
    \includegraphics[width=0.99\textwidth]{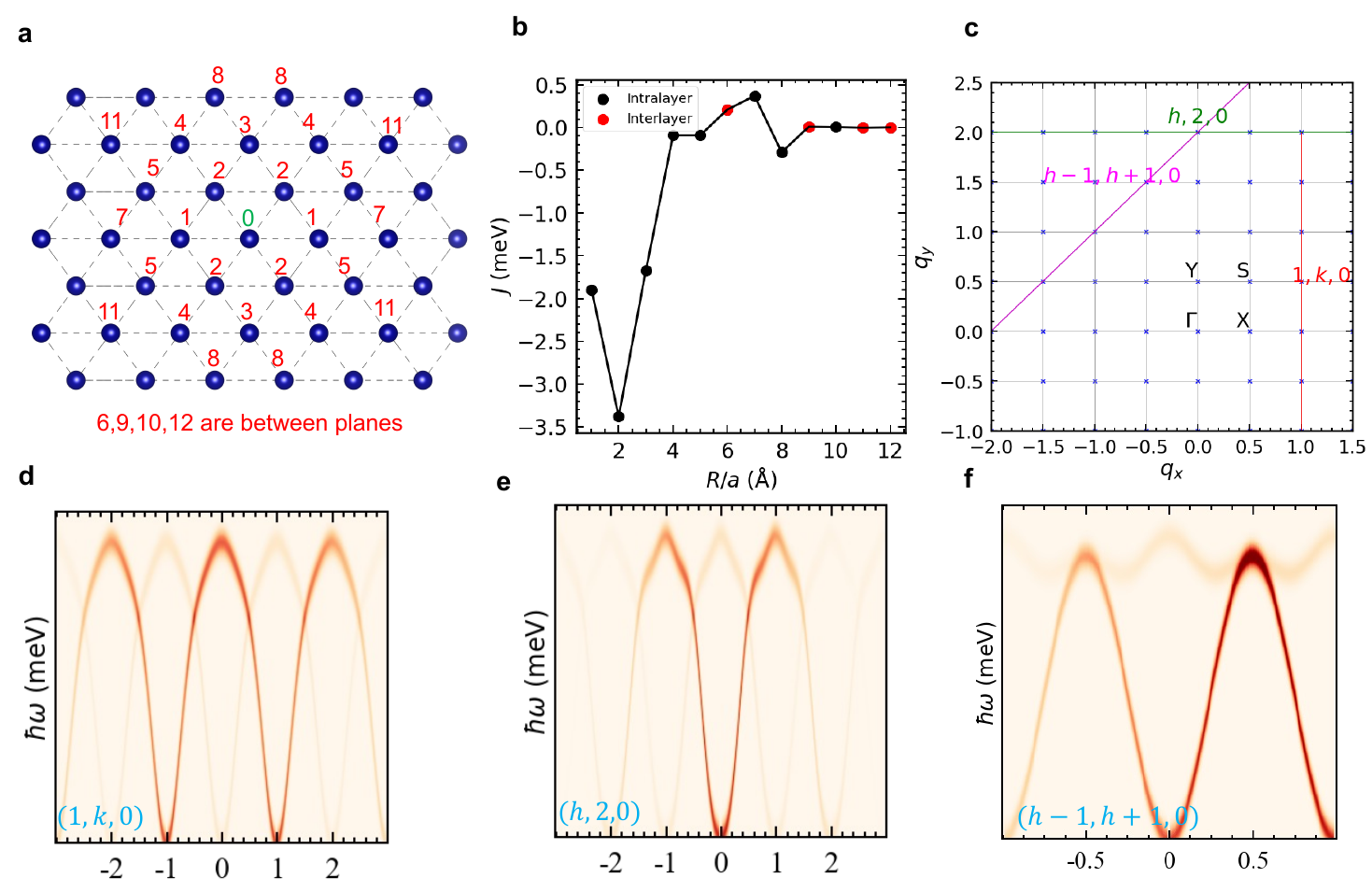}
   \caption{\textbf{Illustration of magnetic properties of bulk CrSBr.} \textbf{a} The top view of bulk CrSBr up to 11th neighbors (6,9,10,12 are between planes). \textbf{b} The values of Heisenberg exchange interaction up to the 12th nearest neighbors where red points are associated to neighbors 6, 9, 10, and 12, which are situated in the adjacent plane to atom 0. \textbf{c} Top view of the first Brillouin zone of bulk CrSBr. \textbf{d-f} Spin-wave dispersion curves for three different wavevector paths: \( (1, k, 0) \), \( (h, 2, 0) \), and \( (h - 1, h + 1, 0) \).}
\end{figure}

\begin{figure}[H]
\centering
   \includegraphics[width=\textwidth]{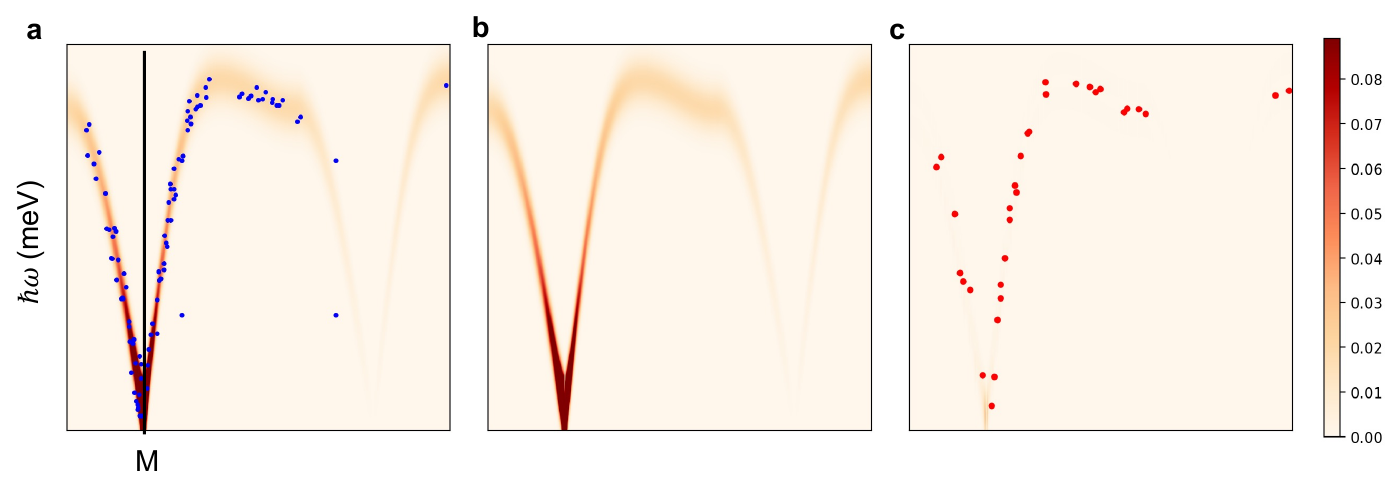}
   \caption{\textbf{Magnon spectrum fitting in La$_2$CuO$_4$ if we have portion information around the high symmetry point M.} \textbf{a} The target magnon spectrum (orange curve) among the three q-paths using the DFT calculated parameters. The blue points are the active sampling points from previous iterations that were suggested by the sBO. \textbf{b} The fBO fitted the magnon spectrum among the three q-paths using only the information from blue sample points. \textbf{c} The absolute intensity deviation between the standard function and fBO’s prediction, and the red points denote the samples to be measured for the next round suggested by sBO. The magnitudes of intensity are described in the color bar.}
    \label{fig-1}
\end{figure}